%% file: beyondT1.tex
\documentclass[aps,pra,reprint,superscriptaddress]{revtex4-1}

\usepackage{amsmath}
\usepackage{amssymb}
\usepackage{graphicx}
\usepackage{textcomp}

\include{macros}
\begin{document}


\title{Nonvolatile quantum memory enables sensor unlimited nanoscale spectroscopy of finite quantum systems \\}


\author{Matthias Pfender}
\email[]{m.pfender@physik.uni-stuttgart.de}
\author{Nabeel Aslam}
\affiliation{3. Physikalisches Institut, University of Stuttgart, Pfaffenwaldring 57, 70569 Stuttgart, Germany}
\author{Hitoshi Sumiya}
\affiliation{Sumitomo Electric Industries Ltd., Itami, 664-0016, Japan}
\author{Shinobu Onoda}
\affiliation{Takasaki Advanced Radiation Research Institute, National Institutes for Quantum and Radiological Science and Technology, Takasaki, 370-1292, Japan}
\author{Philipp Neumann}
\email[]{p.neumann@physik.uni-stuttgart.de}
\affiliation{3. Physikalisches Institut, University of Stuttgart, Pfaffenwaldring 57, 70569 Stuttgart, Germany}
\author{Junichi Isoya}
\affiliation{Research Center for Knowledge Communities, University of Tsukuba, Tsukuba, 305-8550 Japan}
\author{Carlos Meriles}
\affiliation{Department of Physics, CUNY—City College of New York, 160 Convent Avenue, New York, NY 10031, USA}
\author{J\"org Wrachtrup}
\affiliation{3. Physikalisches Institut, University of Stuttgart, Pfaffenwaldring 57, 70569 Stuttgart, Germany}


\date{\today}

\begin{abstract}
In nanoscale metrology applications, measurements are commonly limited by the performance of the sensor.
Here we show that in nanoscale nuclear magnetic resonance (NMR) spectroscopy measurements using single nitrogen-vacancy (NV) centers in diamond, the NV sensor electron spin limits spectral resolution down to a few hundred Hz, which constraints the characterization and coherent control of finite spin systems, and furthermore, is insufficient for high resolution NMR spectroscopy aiming at single molecule recognition and structure analysis of the latter.
To overcome the limitation, we support an NV electron spin sensor with a nuclear spin qubit acting as quantum and classical memory allowing for intermediate nonvolatile storage of metrology information, while suppressing the deleterious back-action of the sensor onto the target system under investigation.
We demonstrate quantum and classical memory lifetimes of $8\,$ms and $4\,$minutes respectively under ambient conditions.
Furthermore, we design and test measurement and decoupling protocols, which exploit such memory qubits efficiently.
Using our hybrid quantum-classical sensor device, we achieve high resolution NMR spectra with linewidths of single spins down to $13\,$Hz.
Our work is therefore a prerequisite for high resolution NMR spectroscopy on nanoscopic quantum systems down to the single level.
\end{abstract}



\maketitle

\section{Introduction}
In the last few years, the nitrogen-vacancy center in diamond has emerged as an exceptional nanoscale quantum sensor, incorporating a single $S=1$ electron spin that can be initialized and read out optically, even at ambient temperatures \cite{doherty_nitrogen-vacancy_2013}.
It constitutes a small quantum processor \cite{dutt_quantum_2007,neumann_multipartite_2008,waldherr_quantum_2014} and an optically accessible, nanoscopic sensor for magnetic and electric fields, temperature and pressure \cite{balasubramanian_nanoscale_2008-1,maze_nanoscale_2008,dolde_nanoscale_2014-2,neumann_high-precision_2013,kucsko_nanometre-scale_2013,doherty_electronic_2014}.
In particular it has been demonstrated that a single NV center can detect proximal nuclear spins inside and outside of the diamond lattice. 
For instance, individual \Ciso\ spins inside and \Siso\ spins outside of the diamond have been detected \cite{taminiau_detection_2012,zaiser_enhancing_2016-1,muller_nuclear_2014}.
Furthermore, proton spin ensembles from a nanoscopic volume can be detected \cite{staudacher_nuclear_2013,mamin_nanoscale_2013,ajoy_atomic-scale_2014} and distinguished from fluorine or silicon nuclear spins via their distinct gyromagnetic ratios \cite{rugar_proton_2015,devience_nanoscale_2015,haberle_nanoscale_2015,muller_nuclear_2014}.
While strongly coupled nuclear spins such as the NV center's intrinsic nitrogen and \Ciso\ are excellent candidates for quantum bits \cite{dutt_quantum_2007,waldherr_quantum_2014}, weaker coupled spins are rather regarded as target spins that can be characterized but lack full individual qubit control.
Nevertheless, there exist quantum simulator proposals based on tailored ensembles of weakly coupled nuclear spins close to an NV center \cite{cai_large-scale_2013}.
While NV centers facilitate target spin identification with spectral linewidths of several hundred Hz \cite{kong_towards_2015,zaiser_enhancing_2016-1}, they are yet unable to identify individual molecules or their structure via their unique NMR fingerprint as in NMR spectroscopy (e.g. chemical shift, J-coupling, few Hz).
\\
\indent
The limiting factor for nanoscale NMR spectroscopy resolution obtained by an NV electron spin sensor is the $T_1$ relaxation of the sensor itself.
The effect is twofold.
First, the sensor spin $T_1$ relaxation process establishes a decay channel between its environment and the target and thus limits the NMR linewidth of the latter.
Second, metrology information about the target spin is irreversibly lost beyond the $T_1$ time, which limits interrogation time and hence spectral resolution of the sensing device.
As a solution, we demonstrate the application of a nuclear spin quantum memory for intermediate storage of quantum and classical metrology information.
The nonvolatile memory allows dissipative decoupling of target spins from the sensor and thus closing the decay channel.
Hence, we retain intrinsic evolution processes of the target spins and metrology information far beyond total decay of the sensor spin, which allows for Hz spectral resolution of the target spin resonances.
\begin{figure*}[t!]
	\begin{center}
	\includegraphics[width=\textwidth]{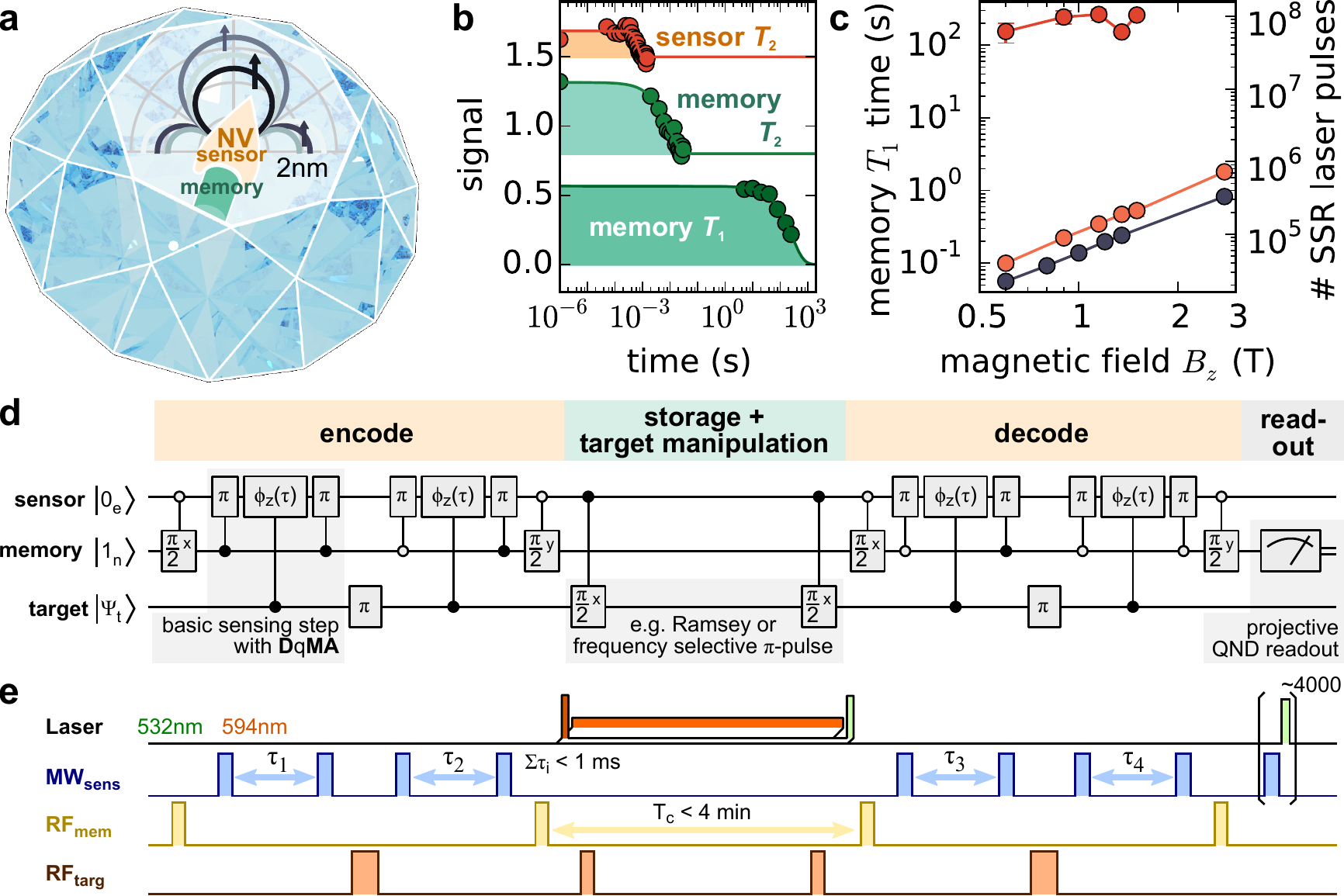}
	\caption{
		\textbf{Introduction to the combined sensor-memory spin system.}
		\textbf{(a)} Schematic representation of our hybrid sensor-memory pair realized by the electron and \Niso\ nuclear spins of an NV defect center in diamond.
		The inset polar plot sketches the potential locations of \Ciso\ target nuclear spins detected in this work.
		\textbf{(b)} \Tonesensor\ decay of the sensor spin (orange), \Ttwomemory\ and \Tonememory\ of the memory spin (upper and lower green line, respectively).
		The transverse relaxation time \Ttwomemory\ of the memory is limited by the longitudinal relaxation of the sensor to about $9$\,ms, while the longitudinal relaxation is orders of magnitude longer (\Tonememory$\approx 260$\,s).
		Therefore, we utilize the memory spin expectation value $\left<I_z\right>$ as a classical memory.
		\textbf{(c)} Shows the \Tonememory-scaling with the external magnetic field for three cases:
		The NV center resides in the negative (red line), the neutral charge state (orange line) and during continuous readout of the memory spin (gray line).
		\textbf{(d)} The wire diagram of the measurement sequence used in this work is subdivided into four different tasks: Encoding; storage and manipulation; decoding and readout.
		During encoding, the sensor writes a phase, which is conditional on the state of the target spin, onto a memory superpositions state.
		This process is efficiently performed by entangling sensor and memory with \cMnotS-gates granting the sensor direct quantum memory access (DqMA).
		Between subsequent DqMA steps we flip the target spins and invert the sign with which DqMA adds phase to the memory.
		The latter part is realized by properly choosing the conditions of the displayed \cMnotS\ gates.
		This protects against quasi static magnetic field noise (see app. \ref{sec:gate_composition},\ref{sec:filter_functions}).
		RF $\pi/2$ pulses with proper phases translate between quantum (i.e. $\ket{0}+e^{i\phi}\ket{1}$) and classical (i.e. $\left<I_z\right>=\sin\phi$) information storage on the memory.
		During the central storage interval we change the target spin states with high spectral resolution, either by an RF $\pi$-pulse or a Ramsey interferometry measurement on the target spins.
		During decoding, the current target spin state is correlated with its encoded initial state, and the result remains as expectation value $\left<I_z\right>$ on the memory.
		We measure the final memory state via single shot readout \cite{neumann_single-shot_2010}.
		\textbf{(e)} Pulse scheme representation of the measurement sequence. 
		Four distinct phase accumulation times $\tau_1$ to $\tau_4$ result in the total sensing time $\tau=\sum_i \tau_i$.
		During the central storage interval \Tcorr, the sensor spin can be continuously repolarized into the \ket{0}, or ionized to the neutral charge state \NVz by appropriate laser illumination.
		\label{fig:1}
	}
	\end{center}
\end{figure*}

\section{Results}
The transverse relaxation time \Ttwostar\ limits the observation of a spin's free precession, and therefore the frequency resolution.
However, since the relaxation time of our sensor is typically shorter than that of the detected target spins, the use of a longer living memory is needed to reach the \ttwostar\ limit.
The longitudinal relaxation time of the NV center sensor spin's expectation value $\left<S_z\right>$ of $\Tonesensor\approx 6\,$ms is usually one to three orders of magnitude longer than its transverse relaxation time \Ttwosensor\ (see Fig.~\ref{fig:1}b).
Therefore, the spin expectation value $\left<S_z\right>$ of the sensor can be used as an intermediate classical memory to improve frequency resolution to about 100\,Hz in a classical correlation measurement \cite{laraoui_high-resolution_2013}.
However, since residual phase information is lost during classical storage, the signal amplitude is decreased by a factor of $\frac{1}{2}$ on average \cite{zaiser_enhancing_2016-1}.
\\
\indent
Classical correlation spectroscopy measurements consist of two phase accumulation times $\tau<\Ttwosensor$, separated by a correlation time $\Tcorr<\Tonesensor$ 
\cite{mims_pulsed_1965,laraoui_high-resolution_2013-2,zaiser_enhancing_2016-1}.
The initial state of the target system is encoded in the first accumulated phase and further stored on the memory.
During \Tcorr\ the target may be manipulated deliberately before the encoded initial state is decoded by the final target state in the second phase accumulation period yielding the correlation result.
It has recently been shown, that for these kind of correlation measurements, the nitrogen nuclear spin, present in every NV center, can be used as a quantum memory \cite{zaiser_enhancing_2016-1}.
Its strong hyperfine coupling to the NV sensor spin (\Nis: $3.03\,$MHz, \Niso: $-2.16\,$MHz) grants fast memory access and therefore does not noticeably shorten the available coherence time \Ttwosensor.
Furthermore, the memory's transverse relaxation time \Ttwomemory\ is only limited by the longitudinal relaxation of the sensor, resulting in quantum storage times on the same order as the lifetime \Tonesensor\ of the sensor spin's polarization (see Fig.~\ref{fig:1}b).
Additionally, owing to the quantum nature of the spin, the full quantum information can be stored, resulting in a higher signal contrast.
Finally, the nitrogen memory spin can be read out in a single shot with high-fidelity \cite{neumann_single-shot_2010}.
Another possibility is to search for an NV center with a weakly coupled \Ciso\ spin.
It has been shown, that a few kHz coupled \Ciso\ memory can keep quantum information on the order of seconds \cite{maurer_room-temperature_2012}.
However, memory access is quite slow and would use up almost the total sensor coherence time \Ttwosensor.
\\
\indent
In this work, we significantly extend the previously reported correlation times and therefore enter a new regime of spectral resolution beyond the \Tonesensor\ limit.
To this end, we analyze and tailor the filter function of the detection scheme, we provide a nonvolatile memory and develop and compare complementary decoupling techniques for the target spins to demonstrate high resolution NMR spectroscopy on individual \Ciso\ target nuclear spins within an isotopically purified diamond crystal ($\left[ ^{12}\mathrm{C} \right] =0.99995$).
As a main point of this work, we utilize the expectation value $\left\langle I_z \right\rangle$ of the nitrogen nuclear spin as intermediate classical memory, which reaches minutes-long storage times at room temperature (see Fig.~\ref{fig:1}b).
We further show that under certain conditions either orange ($594\,$nm) laser illumination of the NV center or switching its charge state can serve as a dissipative decoupling technique for the target spins.
\\
\indent
For efficient encoding and decoding of metrology data on the memory, the sensor has direct quantum memory access (DqMA) via mutual entanglement \cite{zaiser_enhancing_2016} avoiding long-lasting SWAP-gates (see figure \ref{fig:1}d).
To this end, a superposition state of the memory is entangled with the sensor by use of a \cMnotS\ gate, causing the sensor-memory system to acquire a quantum phase dependent on an external magnetic field (e.g. the Overhauser field of a target spin).
A second \cMnotS\ gate disentangles the two spins, leaving the acquired phase on the memory superposition state.
By carefully choosing the control conditions of these gates, the sign of the phase as well as the resulting sensor state can be chosen (see Appendix \ref{sec:gate_composition}).
During the correlation and storage interval \Tcorr, we manipulate target nuclear spins with radiofrequency (RF) pulses without influencing the information on the memory.
Target spins that are flipped during \Tcorr\ contribute maximally to the correlation result in the decoding process, while the effect of quasi-static magnetic field noise is filtered out (see Appendix \ref{sec:filter_functions}).
Furthermore, arbitrary NMR pulse sequences can be applied to induce flips of target nuclear spins during the central RF manipulation period.
In this work, we either perform an ordinary, frequency selective $\pi$-pulse or a Ramsey sequence.
\\
\indent
\begin{figure}[t!]
	\begin{center}
	\includegraphics[width=\columnwidth]{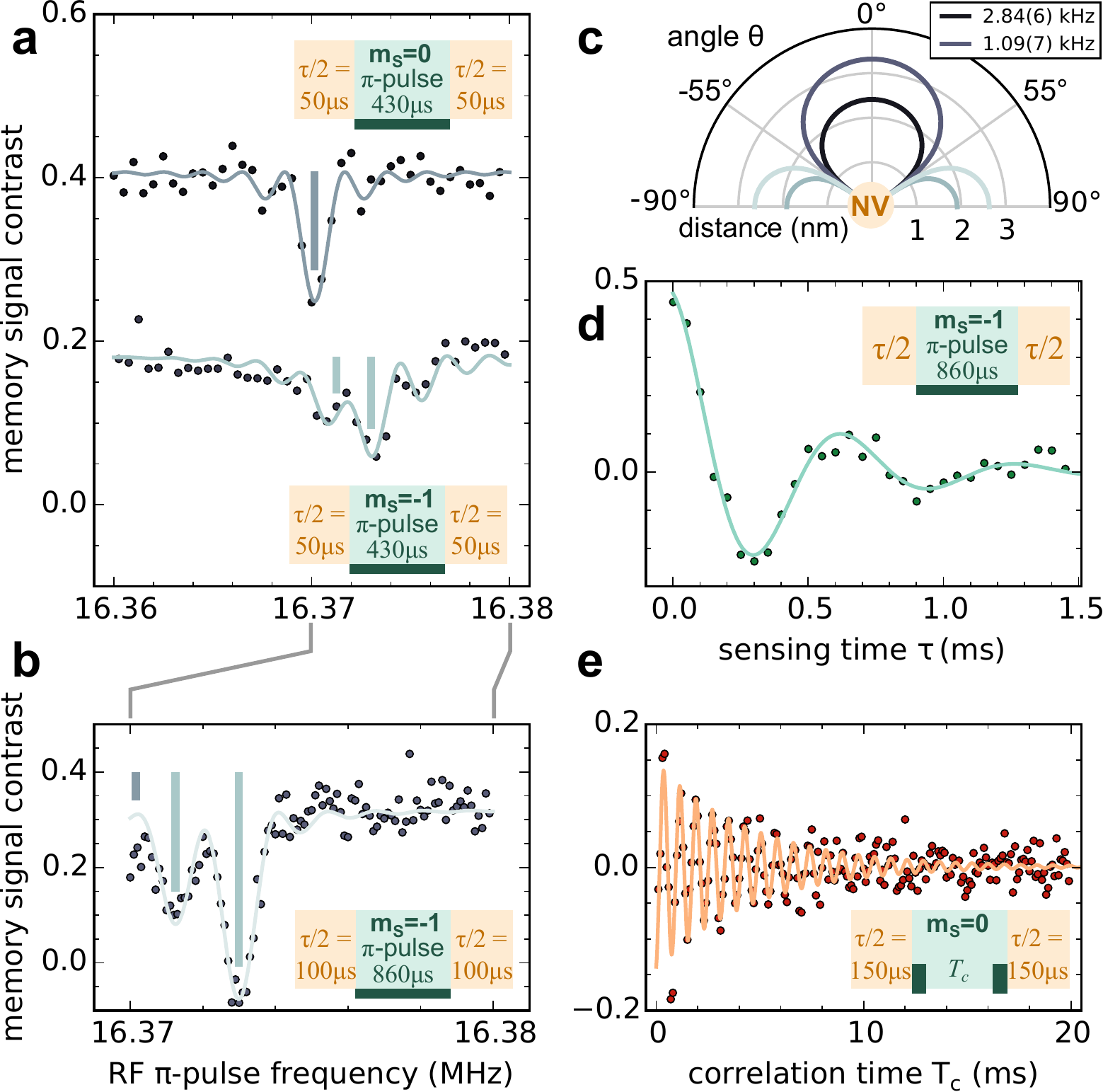}
	\caption{
		\textbf{Spectroscopy of distant, weakly coupled \Ciso\ spins limited by sensor dissipation.} 
		\textbf{(a)} Spectra of the \Ciso\ target spins surrounding the used NV center.
		For the upper (lower) \Ciso\ spectrum, the sensor spin is in its non-magnetic $\ket{m_S=0}$ (magnetic $\ket{-1}$) state during the quantum storage interval.
		Hence, the upper curve shows the bare Larmor frequency common to all \Ciso\ nuclear spins in the interaction range, whereas the lower curve shows additional hyperfine coupling offsets \Azz\ for two distinct nuclei.
		The sensing time is set to $\tau=100\,\mu$s (orange parts of insets) and the RF $\pi$-pulse (dark green part of insets), which flips the \Ciso\ spins during the storage interval, is set to $T_c=430\,\mu$s (light green part of insets).
		\textbf{(b)}
		Increased sensing and storage time ($\tau=200\,\mu$s and $T_c=860\,\mu$s, respectively) generate a close-up spectrum of individual \Ciso\ spins. 
		Resonance frequencies are marked with vertical lines.
		\textbf{(c)}
		Possible angles $\theta$ and distances $d$ of the two \Ciso\ spins relative to the NV center.
		The functions are $d \propto ( \Azz^{-1} | 3 \cos^2\theta-1 | ) ^{1/3}$.
		For the identified hyperfine interactions \Azz, $3 \cos^2\theta-1$ is positive (dark line parts for $\theta=-54.7^{\circ} \ldots 54.7^{\circ}$).
		\textbf{(d)}
		Variation of the overall sensing time $\tau$ for an RF $\pi$-pulse selective on the strongest coupled \Ciso\ spin ($\nu_{\mathrm{RF}}\approx 16.373\,$MHz, $T_c=860\,\mu$s) reveals an oscillation, which confirms the coupling strength of $2.8\,$kHz.
		\textbf{(e)}
		For a sensing time of $\tau=300\,\mu$s the memory state is almost maximally correlated with the flip of the $2.8\,$kHz \Ciso\ spin.
		During storage we perform a Ramsey oscillation on the \Ciso\ spin (inset shows two dark green $\pi/2$-pulses), which is converted into a memory signal and decays with the memory lifetime, which is limited by dissipation due to the sensor.
		\label{fig:2}
	}
	\end{center}
\end{figure}
Fig.~\ref{fig:2} a, b show exemplary NMR spectra of weakly coupled \Ciso\ target spins obtained with our measurement sequence as follows.
We encode and decode an individual target spin state efficiently, if the total sensing time $\tau$ is around $\Azz^{-1}$ \cite{zaiser_enhancing_2016-1}, where \Azz\ is the expected hyperfine coupling between sensor and target spin 
($\tau=100$ and $200\,\mu$s in Fig.~\ref{fig:2}a and b, respectively).
The inset in Fig.~\ref{fig:2}~a,~b shows a simplified illustration of the measurement scheme comprising of the coding and storage intervals (compare Figs. \ref{fig:1} and \ref{fig:app_gates}).
During the whole central storage and correlation interval \Tcorr\ we apply a constant-frequency $\pi$-pulse ($430$ and $860\,\mu$s in Fig.~\ref{fig:2}a and b), which is selective in a narrow spectral range of about $\pm1$ and $\pm0.5\,$kHz.
We acquire a target spin spectrum by stepping the $\pi$-pulse frequency.
While the upper spectrum in Fig.~\ref{fig:2}a reveals one resonance at the bare \Ciso\ Larmor frequency, the lower spectrum and its zoom-in in Fig.~\ref{fig:2}b show multiple target spins shifted by their individual hyperfine coupling \Azz.
To this end, we have set the sensor into its non-magnetic $\ket{m_S=0}$ or magnetic $\ket{m_S=1}$ state during \Tcorr, respectively.
The latter case switches on coupling and therefore the possibility to manipulate individual target spins conditional on their coupling strength.
Possible locations of the two spectrally resolved \Ciso\ spins with respect to the NV center in the diamond lattice are displayed in Fig.~\ref{fig:2}c.
\\
\indent
Following the results of Fig.~\ref{fig:2}b we set the RF $\pi$-pulse frequency resonant to the $2.8\,$kHz coupled \Ciso\ spin, with a duration of $860\,\mu$s.
When changing the total sensing time $\tau$, the measurement signal oscillates with the coupling strength (see Fig.~\ref{fig:2}d), hinting towards a single target spin.
We establish maximum correlation of sensor and target \Ciso\ spin, and therefore prepare a strong, projective measurement of the target spin, by setting $\tau=300\,\mu$s \cite{ralph_quantum_2006}.
\\
\indent
With these settings, we perform a Ramsey experiment selectively on the $2.8\,$kHz coupled target spin (see Fig.~\ref{fig:1}b).
Fig.~\ref{fig:2}e shows the resulting Ramsey oscillation of the \Ciso\ spin, which decays exponentially with a time constant of $4.5\,$ms, corresponding to a linewidth of 71\,Hz.
In contrast, the expected decay constant of a single \Ciso\ spin in such an isotopically purified diamond is expected to be less than $1\,$Hz (the coupling of two \Ciso\ spins at average distance is around 0.07\,Hz).
The main reason for comparably fast signal decay are the probabilistic sensor spin flips on a timescale \Tonesensor\ leading to fluctuating magnetic fields and hence resonance frequency shifts for both, memory and target spin.
Thus, on the one hand, the quantum state stored on the memory dephases and the metrology information gets lost with increasing \Tcorr.
On the other hand, also the target spin dephases and thus does not exhibit its intrinsic \ttwostar\ time (see Appendix~\ref{app:relaxation_theory}).
\\
\indent
Since we cannot decrease the sensor's dissipation rate $1/\Tonesensor$ at room temperature \cite{jarmola_temperature-_2012}, we need to mitigate its effect on the stored metrology information and on the target spins.
The sensor's dissipation affects the memory by dephasing the stored information but not by scrambling its expectation value $\left<I_z\right>$.
Therefore, a natural choice for less volatile information storage is the memory spin's expectation value $\left<I_z\right>$.
When measuring the \Tonememory\ time of the memory spin for magnetic fields in the range of $0.6$ to $1.5\,$T, we achieve values above 100 seconds, which increase up to $240\,$s for the highest measured field strength (see Fig.~\ref{fig:1}b, c and \ref{fig:app_C_1}a).
Lifetimes that long are due to the lack of noise sources at the memory's resonance frequency (e.g. no bath of other nitrogen nuclear spins) and due to a reduction of flip-flop processes with the sensor spin \cite{neumann_single-shot_2010}.
Since the initial metrology information is encoded in the phase of the quantum memory (i.e. $\ee^{\ii \phi}$), we transfer part of the sensing information on the expectation value $\left<I_z\right>$ of the memory (i.e. $\sin \phi$) utilizing a $\pi/2$-pulse on the memory spin, prior and after the central correlation time \Tcorr\ (cf. Fig.~\ref{fig:1}d).
\\
\indent
In addition to a protected memory, we need to prevent decoherence of the target spins.
This can be achieved by increased dissipation of the sensor (dissipative decoupling), either via continuous optical pumping into the $\ket{0}$ eigenstate, or by exploiting the increased electron spin dissipation rate in the NV center's neutral charge state \cite{waldherr_distinguishing_2012}.
Analogous to motional averaging in liquid-state NMR experiments, the effect of the coupling between sensor and target spins is averaged out and the coherence time of the target spins is prolonged (compare \cite{maurer_room-temperature_2012,li_motional_2013,borregaard_scalable_2016}).
Therefore, we seek to increase the sensor dissipation rate $\Gamma$ much beyond the sensor-target coupling strength $\Gamma\gg\Azz$ for the target spin linewidth $\delta\nu=(\pi \ttwostar)^{-1}$ to scale as
\begin{equation}
\delta\nu\propto\Azz^2\Gamma^{-1}
\label{eq:motional_average}
\end{equation}
(see appendix~\ref{sec:master_equation}).
\\
\indent
\begin{figure}[t!]
	\begin{center}
	\includegraphics[width=\columnwidth]{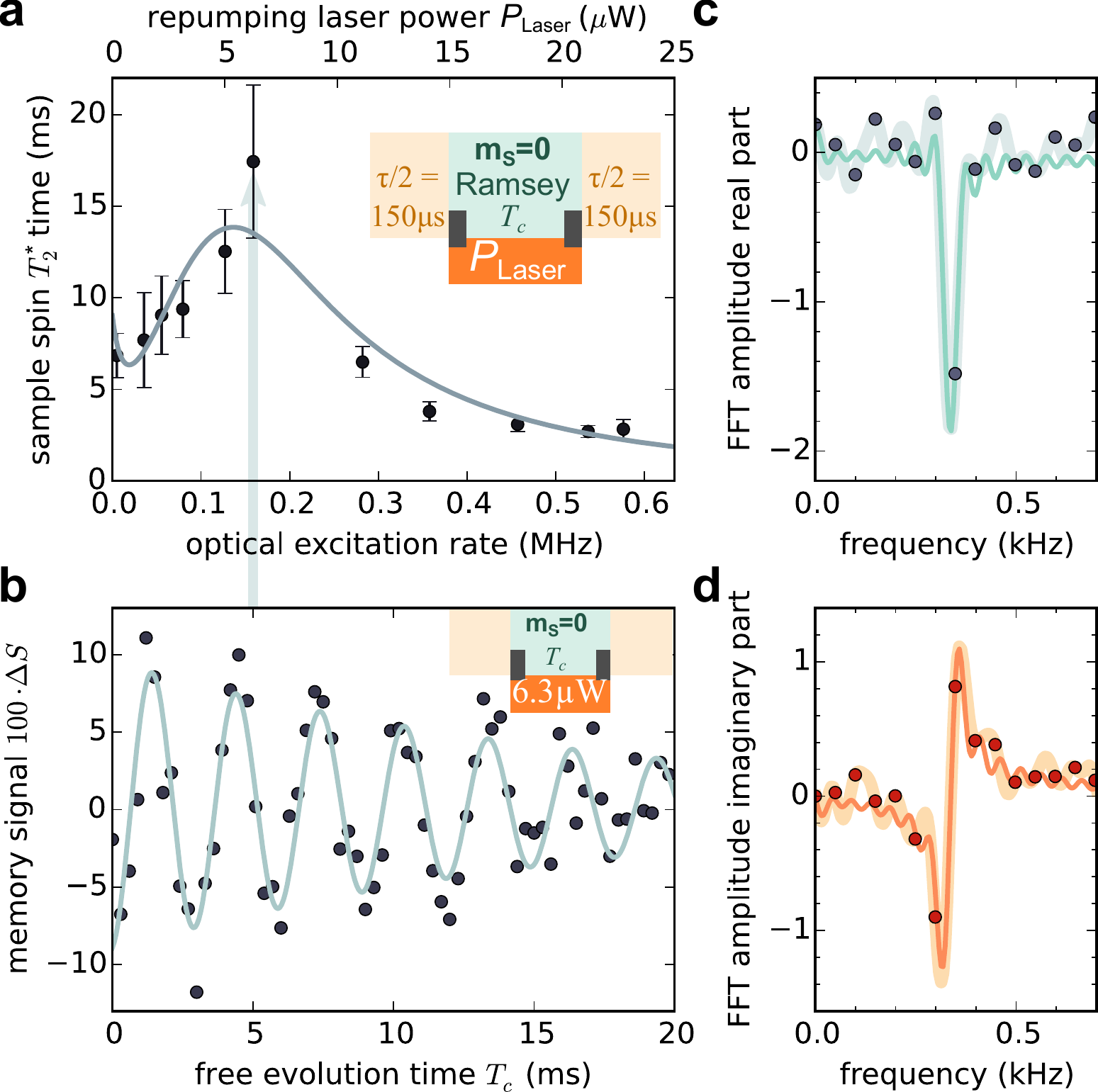}
	\caption{
		\textbf{Dissipative decoupling of target spins during classical storage on quantum memory.} 
		\textbf{(a)} During storage (green background) the sensor spin flips stochastically into a magnetic state $\ket{\pm1}$ and an orange repumping Laser ($594\,$nm) excites the NV and repolarizes the sensor spin predominantly into the non-magnetic state $\ket{0}$.
		We perform Ramsey measurements on the 2.8\,kHz coupled target spin from figure \ref{fig:2} and record the \Ttwostar\ time depending on the applied optical excitation rate. 
		The resulting lifetimes are fitted with simulated coherence lifetimes, using only the conversion from applied laser power to excitation rate, the ionization probability and the probability for the $\ket{\pm1}$ excited state to decay via the metastable state into the $\ket{0}$ ground state (see appendix \ref{sec:increased-dissipation-by-illumination}).
		We observe three distinct regions (see appendix \ref{sec:increased-dissipation-by-illumination}).
		The first decrease is induced by a slow excitation and depolarization into the $\ket{\pm 1}$ spin state.
		As soon as the excitation rate exceeds the coupling strength, the coherence time starts to increase.
		At a certain excitation rate, the sensor is ionized before the measurement can be concluded, resulting in loss of signal.
		\textbf{(b)}
		The target spin's ($2.8\,$kHz coupled \Ciso\ spin from fig.~\ref{fig:2}) longest coherence time \Ttwostar\ for an applied laser power of $6.3\,\mu\text{W}$.
		The straight line is a fit of a decaying cosine oscillation, with a decay time \Ttwostar\ of $17.4\pm4.2$\,ms.
		\textbf{(c,d)}
		Show the real (imaginary) part of the Fourier transformed signal of \textbf{(b)}. While the dark gray (orange) spots represent the raw transform, the light gray (orange) line represents the transformation of the zero-filled input signal.
		The green (orange) line show a fit, where the real- as well as the imaginary part of the Fourier transform are fitted in a combined manner.
		We obtain a linewidth of $18.3\pm4.3$\,Hz.
		\label{fig:3}
	}
	\end{center}
\end{figure}
In the case of continuous optical sensor spin initialization, the dissipation rate is related to the excitation rate as $\Gamma\propto\gamma_\mathrm{exc}$.
Hence, a higher excitation rate is beneficial for the dissipative decoupling effect on the target nuclear spins \cite{maurer_room-temperature_2012}.
In addition, the sensor spin initialization probability into $\ket{0}$ of $\approx98\,\%$ \cite{waldherr_dark_2011} reduces the average coupling \Azz\ and hence, the pre-factor in equation (\ref{eq:motional_average}).
However, high excitation rates also cause decay of the memory qubit spin expectation value $\left<I_z\right>$, both by ionization of the NV center to its neutral charge state with rate $\gamma_\mathrm{ion}\propto\gamma_\mathrm{exc}^2$ and by higher occupation probability of the electronic excited state \cite{aslam_photo-induced_2013,fuchs_excited-state_2008,neumann_excited-state_2009,neumann_single-shot_2010,pfender_single-spin_2014}.
By carefully choosing the excitation power at a wavelength of 594\,nm \cite{aslam_photo-induced_2013,pfender_single-spin_2014}, one can set the ionization limit on the linewidth $\delta\nu$ equal to the dissipative decoupling limit according to equation (\ref{eq:motional_average}) (see appendix~\ref{sec:master_equation}).
We measure the Ramsey oscillation decay of the $2.8\,$kHz coupled \Ciso\ target spin for varying optical excitation power and reveal an enhancement of the target spin coherence time by almost a factor of four up to $\ttwostar=17.4\pm4.2$\,ms, at a repumping laser power of 6.3\,\textmu W (see Fig.~\ref{fig:3}b).
The corresponding Fourier transformation reveals a sharp peak with a FWHM of $18.3\pm 4.3$\,Hz.
\\
\indent
Further increasing the excitation rate causes the NV center to ionize faster, resulting in a faster loss of signal and an apparently shorter \ttwostar\ of the target spin (see Fig.~\ref{fig:3}a).
However, the latter loss of signal neither implies decay of the classical information stored on the memory nor faster instrinsic dephasing of the target spin.
It is due to the lack of memory access.
In order to restore access, we recover the negative charge state after the correlation time \Tcorr\ and before the final decoding part of the sequence by a short green (532 nm) laser pulse (see Figs.~\ref{fig:4}c and \ref{fig:1}d).
It turns out, that the uncharged state of the NV center can also be used for decoupling of target spins, which constitutes our second dissipative decoupling approach.
Owing to the fast dissipation rate $\Gamma_{\text{NV}^0}$ of the electron spin in \NVz\ state \cite{waldherr_dark_2011}, no laser excitation during \Tcorr\ is required.
Instead, intrinsic fast flips of the electron spin lead to an effective decoupling for small couplings $\Azz\ll \Gamma_{\text{NV}}^0$.
\\
\indent
\begin{figure*}[t!]
	\begin{center}
	\includegraphics[width=\textwidth]{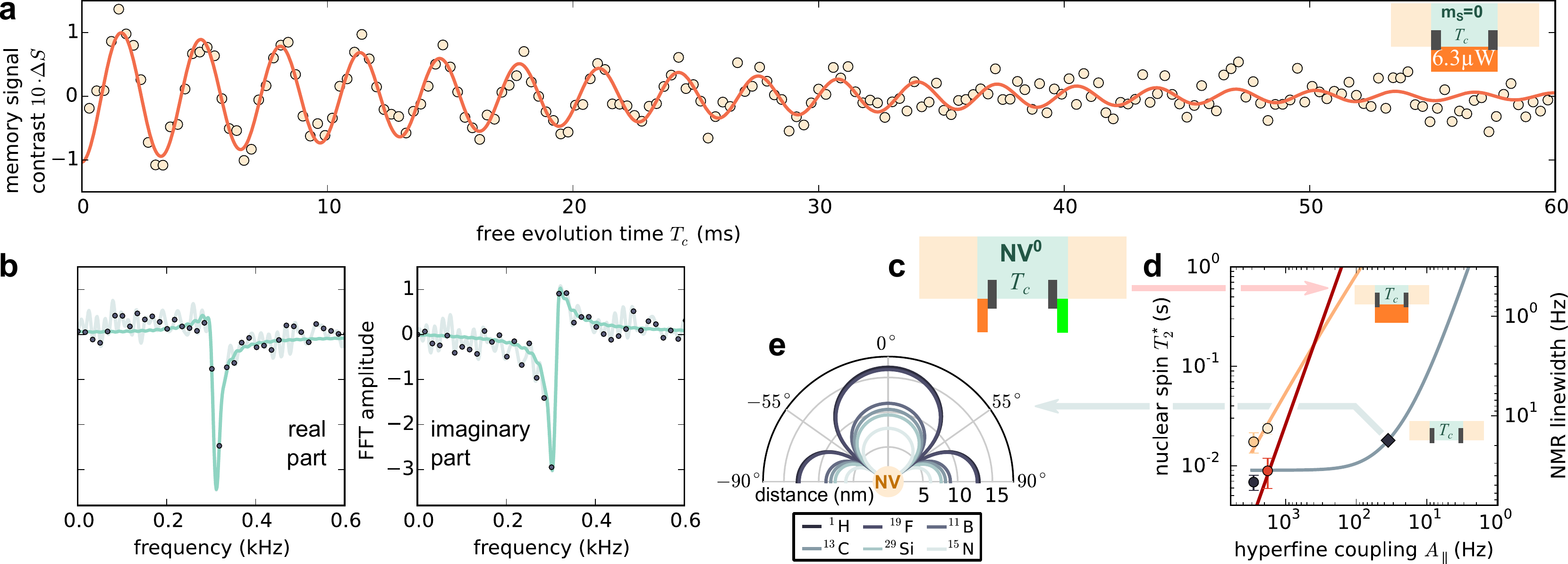}
	\caption{
		\textbf{Scaling of decoupling techniques for improved spectral resolution.}
		\textbf{(a)} Ramsey oscillations of a different NV with a $-1.8\,$kHz coupled \Ciso\ spin during constant sensor spin repumping. The decay constant of the oscillation is $23.8\pm2.9\,$ms.
		\textbf{(b)} shows the Fourier transform of the signal. Analogous to Fig. \ref{fig:3}c, d the dark gray points are the transform of the unprocessed measurement data, the light gray line is the transform of the zero-filled signal.
		The green line is the simultaneous fit of the real and imaginary part of the transformation.
		The linewidth was determined to be $13.3\pm 1.6$\,Hz.
		\textbf{(c)} Simplified measurement scheme of the target decoherence while the sensor is in its neutral charge state.
		Before applying the first $\pi/2$ pulse on the target spin, the NV center is pumped to the \NVz\ state by applying a $1$\,ms orange laser pulse with a power of $100\,\mu$W.
		The charge state is afterwards recovered by applying a short green laser pulse.
		\textbf{(d)} Simulated coherence times of a target nuclear spin coupled to an NV center, when using one of the aforementioned decoupling methods.
		The gray curve is for the plain \NVm\, the red for the \NVz\ case.
		The orange line shows the coherence time, when using the optimal repolarization rate for every coupling (see appendix \ref{sec:increased-dissipation-by-illumination}).
		All measured coherence times of this work are displayed as correspondingly color-coded circles.
		For 1\,kHz coupling, the resulting coherence times are: 9\,ms for \NVm, 26\,ms for \NVz and 54\,ms for the repumping case.
		The gray diamond marks the coupling strength at which the plain \NVm\ exerts a dissipative decoupling effect, which doubles the target spins coherence time.
		In \textbf{(e)}, the possible location of equi-coupled nuclear spins of different spin species are shown (see diamond in panel \textbf{d}).
		Therefore, the dissipative decoupling technique developed here is of significant importance when performing NV high resolution NMR spectroscopy, especially when working with small ensembles of target spins.
		\label{fig:4}
	}
	\end{center}
\end{figure*}
We benchmark the two dissipative decoupling techniques on a different NV center, with a \Ciso\ nuclear spin weaker coupled than the previous one (i.e. $-1.8\,$kHz instead of $2.8\,$kHz, see Fig.~\ref{fig:4}a-c). 
When utilizing the continuous optical repolarization method, we obtain a Ramsey oscillation with a decay constant of $23.8\pm2.9\,$ms and the Fourier transform reveals a corresponding linewidth of $13.3\pm1.6\,$Hz (see Fig.~\ref{fig:4}b).
For comparison, we measure the linewidth with the NV center being in its neutral charge state (see Fig.~\ref{fig:4}c, d). 
To this end, we switch the NV center during the central correlation interval first into the neutral charge state by a strong orange laser pulse ($100\mu$W, 1\,ms duration), then perform a Ramsey experiment, and finally recover the negative charge state by a green laser pulse and conclude the measurement (see Fig.~\ref{fig:4}e), resulting in a linewidth of $40.8\,$Hz.
Note, that the restoration fidelity of \NVm\ is limited to about 70\% \cite{waldherr_dark_2011}, which directly translates to a reduced signal intensity.
\\
\indent
For the purpose of protecting the coherence of the current $-1.8\,$kHz coupled \Ciso\ spin, the optical repolarization method is clearly superior to the \NVz\ method.
However, when considering the scaling of the expected target spin coherence lifetime with decreasing coupling for both methods, we observe an intercept point at a coupling strength of $\Azz\approx400\,$Hz and a target spin coherence time of $\ttwostar\approx150\,$ms (see Fig.~\ref{fig:4}d).
To this end, we extrapolate the performance of our decoupling techniques for smaller couplings \Azz\ as follows.
Comparing our \NVz\ results with those of Ref.~\onlinecite{waldherr_dark_2011} (i.e. $\left|\Azz\right|=6.06\,$MHz, $\ttwostar\approx 6\,\mu$s), we conclude that in the present case we are well within the motional averaging regime and therefore the coherence time is expected to scale as $\ttwostar\propto\Azz^{-2}\Gamma_{\text{NV}^0}$.
\\
\indent
For the optical repolarization technique, however, we have to set an optimum excitation rate, which decreases with decreasing coupling strength and therefore also $\Gamma_{\text{exc}}$ decreases as $\Gamma_{\text{exc}}\propto\Azz^{2/3}$ (see appendix~\ref{sec:master_equation}).
Hence, we expect a power-law scaling of $\ttwostar\propto\Azz^{-4/3}$.
Given our experimental data we simulate the expected \ttwostar\ for target spins with different couplings (see appendix \ref{app:relaxation_theory}) and get a curve shown in Fig.~\ref{fig:4}d, which agrees with the expected power-law scaling.
\\
\indent
For comparison, we consider an NV center remaining in the negative charge state during the correlation time without any dissipative decoupling technique.
Then, down to sensor-target coupling strengths of about 100\,Hz the target spin coherence time \ttwostar\ would be pinned at 9\,ms and only for even smaller couplings \ttwostar\ would rise.
For example, a target spin coherence time of $18\,$ms is expected for a coupling strength of $\Azz \approx 35\,$Hz (see diamond in fig. \ref{fig:4}d).
Such a coupling appears for a \Ciso-NV distance of up to $10\,$nm or a \proton-NV distance of up to $17\,$nm (see fig. \ref{fig:4}e).

\section{Summary and Conclusion}
In this work we have implemented a hybrid system comprised of an electron spin sensor for magnetic fields and a nuclear spin memory capable of quantum and classical metrology data storage.
It turns out, that the memory's $\left<I_z\right>$ expectation value is very robust and exhibits lifetimes on the order of seconds to minutes for various conditions ranging from no operation at all to continuous laser illumination, continuous storage access and charge state switching of the associated NV center in diamond.
Therefore, we have developed a correlation spectroscopy technique, which exploits the sensor's coupling to magnetic fields and the memory's storage capabilities.
By encoding a fingerprint of the current magnetization of target spins on the memory, the magnetization change induced by a coherent target spin manipulation can be decoded and correlated on timescales up to the \Tonememory\ lifetime.
\\
\indent
The demonstrated sensor-memory approach enables performing high resolution spectroscopy on a few target \Ciso\ spins in diamond.
Here, the long-lived intermediate classical storage increases the spectral resolution far beyond the capabilities of the sensor alone.
However, under these conditions, we start seeing a strong deleterious influence of the sensor spin on the target spins' coherence properties.
We have developed dissipative decoupling techniques that on the one hand preserve the information stored on the memory and on the other hand efficiently reduce the deleterious effect of the sensor on the target spins.
Consequently, we are able to measure nuclear magnetic resonance linewidths of single \Ciso\ spins in diamond of $13.3\,$Hz.
\\
\indent
While we have confirmed their performance for target spins with coupling strengths down to a few kHz, we have also simulated the performance for spins up to the coherent coupling limit of around 300\,Hz and beyond.
We set the latter limit tentatively by the maximum coherence time of the NV sensor spin of up to 3\,ms at room-temperature measured so far \cite{balasubramanian_ultralong_2009}.
For such target spins, dissipative decoupling is expected to improve the coherence time and hence spectral resolution by a factor of about 20.
\\
\indent
Below a coupling strength of 300\,Hz down to about 30\,Hz we enter the weak measurement regime, where the encoded and decoded sensor signal of individual target spins decreases along with a reduced measurement back-action onto the target spins \cite{ralph_quantum_2006}.
However, the dissipative back-action would still deteriorate target spins.
For example, in nanoscale NMR experiments \cite{staudacher_nuclear_2013,mamin_nanoscale_2013,haberle_nanoscale_2015,devience_nanoscale_2015}, where clusters of proton spins at a distance of $\sim10\,$nm to the NV sensor and a corresponding coupling of about 160\,Hz are detected, the dissipative back-action of the sensor on the target spins still limits their coherence times to about 10\,ms.
Hence, in several previous NV-based NMR measurements on external nuclear spins \cite{staudacher_nuclear_2013,mamin_nanoscale_2013,devience_nanoscale_2015,haberle_nanoscale_2015}, the NV sensor itself would have limited the NMR linewidth.
Here, our dissipative decoupling methods would increase the target lifetime by a factor of about 100.
\\
\indent
For even weaker coupled spins, it becomes hardly possible to detect single target spins and we start entering the regime of a classical target system, where both measurement and dissipative back-action become less and less important.
In this classical regime, correlation measurement techniques as described here, which avoid probing and therefore disturbing the targets during the correlation period, are not essential.
The target can rather be measured continuously without any back-action.

\appendix

\section{experimental setup and diamond sample}
\label{app:setup}
Our experiment consists of a home-built confocal microscope with $532\,$nm 
and $594\,$nm excitation lasers, referred to as green and orange lasers, respectively.
The cw lasers can be switched on and off on the timescale of $\sim10\,$ns with an acousto-optical modulator (AOM) and reaches on/off intensity ratios of up to $10^6$.
To completely block illumination lasers, we can additionally flip a beam block into the optical path on timescales of seconds.
The emitted photoluminescence of single NV centers is collected via an oil-immersion objective 
with an NA of 1.35 and detected by a single photon counting detector (avalanche photodiode). 
The sample is positioned inside a $B_z=3\,$T, $B_{x,y}=0.2\,$T superconducting vector magnet, with the diamond's surface normal pointing along the main magnetic field axis \cite{aslam_single_2015-2}.
The magnet is mainly operated at a field of about $1.5\,$T, which shifts the NV sensor spin resonance frequency to about $40\,$GHz.
Spin resonance is detected optically (ODMR) via spin state dependent fluorescence of single NV centers \cite{gruber_scanning_1997}.
Microwave (MW) radiation for hyperfine selective sensor spin manipulation is generated by amplitude modulation (at frequencies of $\sim 1$GHz) of a carrier signal (at $\approx40\,$GHz), 
utilizing a 
double balanced harmonic mixer and an 
arbitrary waveform generator (AWG).
The microwaves are guided through coaxial cables and a coplanar waveguide.
The latter is tapered to around $100\mu$m at the location of the NV for a larger MW field strength.
The RF signal used to manipulate the nuclear spins is created directly by the AWG 
and guided through a $20\,\mu$m copper wire spanned across the diamond perpendicular to the CPW direction.
\\
\indent
The diamond sample is a polished, (111)-oriented slice ($2\,\mathrm{mm}\times 2\,\mathrm{mm}\times 88\,\mu$m) from a high pressure and high temperature (HPHT) diamond crystal ($5.3\,\text{mm}\times 4.7\,\text{mm}\times 2.6\,\text{mm}$).
The original sample is isotopically enriched with a \Cis-concentration of 99.995\%.
The crystal was irradiated by $2\,$MeV electrons at room temperature to the total fluence of $1.3\cdot 10^{11}\,\text{cm}^{-2}$ and annealed at $1000\,^{\circ}\text{C}$ (for 2 hours in vacuum) to create bulk NV centers from intrinsic nitrogen impurities.

\section{Detailed measurement sequences}
\begin{figure}[t!]
	\begin{center}
	\includegraphics[width=\columnwidth]{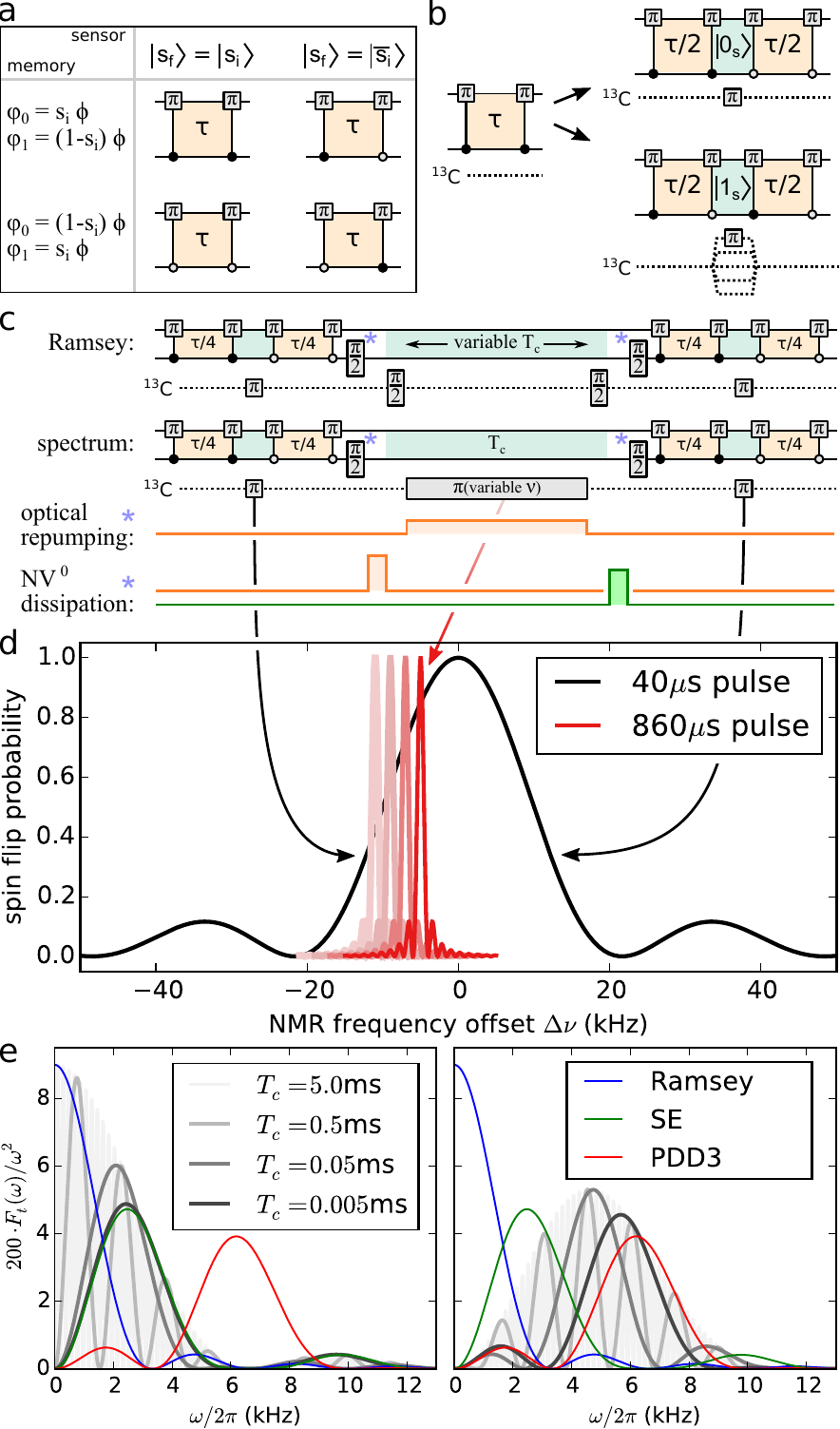}
	\caption{
		\textbf{Design of sensing sequences}.
		\textbf{(a)} shows the four DqMA blocks used to create different sensing tasks.
		Every sequence starts by preparing the sensor spin in an eigenstate and the memory spin in a superposition state.
		These gates have two tasks: Accumulate a phase on one of the components of the memory spin, and either flip or keep the sensor spin state.
		\textbf{(b)} shows the refinement of a Ramsey type sensing gate, so that the accumulated phase only stems from a certain source (here: \Ciso nuclear spins flipped by a RF $\pi$ pulse. This can be done selective, by turning on the hyperfine gradient during the pulse (lower part)).
		In \textbf{(c)}, the sensing sequences used in Fig.~\ref{fig:2} to \ref{fig:4} are shown.
		The outer $\pi$ pulses on the \Ciso\ spins filter the accumulated signal quite broad, while the central, slow $\pi$ pulse results in a high frequency resolution.
		This is shown in \textbf{(d)}.
		The overall signal results from spins that are flipped by all three pulses.
		\textbf{(e)} shows the noise filter functions of the used detection sequences, in comparison to commonly known detection schemes.
		On the left, the sensing gates as shown in \textbf{a} are used, while the right graph incorporates further decoupling, as shown in \textbf{b}.
		\label{fig:app_gates}
	}
	\end{center}
\end{figure}
Here, we explain the applied measurement sequences in more detail.
Therefore, we take a closer look on the individual sensing parts to reveal how magnetic field sensing information is measured and stored on the memory and how it is spectrally filtered to be sensitive mainly to the target spins.

\subsection{Magnetic sensing and storage}
\label{sec:magnetic_sensing_and_storage}
A full measurement sequence consists of two sensing (encoding and decoding) and one intermediate storage interval.
In the experiments performed here, the encoding as well as the decoding interval again is sub-divided into two sensing steps and one intermediate, much shorter, storage step (see Fig.~\ref{fig:app_gates}b, c).
A proper timing of sensing and storing constitutes a dynamical decoupling sequence that renders the sensor-memory-system sensitive to target spins and almost insensitive to other ``noise'' sources.
\\
\indent
During sensing steps, the sensor-memory-system is sensitive to any magnetic field source and thus picks up a corresponding phase $\varphi$.
During storage, the latter sensitivity is effectively switched off and changes to the environment (e.g. the target spins) can be performed without influencing the intermediate sensing result on the memory.
\\
\indent
We utilize entangled states of the sensor and memory qubits for sensing and simultaneous direct quantum memory access during sensing steps (e.g. $\ket{\Psi_\mathrm{s,m}}=e^{\ii\phi_0}\ket{00}+e^{\ii\phi_1}\ket{11}$).
The qubit states $\ket{0}$ and $\ket{1}$ of sensor and memory are related to the magnetic quantum numbers $m_S=0,-1$ and $m_I=0,+1$ for the NV electron and nuclear spin, respectively.
Initial and final sensor-memory quantum states, as well as states during the storage steps are product states $\ket{\Psi_\mathrm{s}}\otimes\ket{\Psi_\mathrm{m}}$ of a sensor Eigenstate $\ket{\Psi_\mathrm{s}}=\ket{0}$, $\ket{1}$ 
and a memory superposition state $\ket{\Psi_\mathrm{m}}$, 
\begin{align}
\label{eq:storage_state}
\ket{\Psi_\mathrm{s}} \otimes \ket{\Psi_\mathrm{m}} &= \ket{\Psi_\mathrm{s}} \otimes \left( e^{\ii\phi_0}\ket{0}+e^{\ii\phi_1}\ket{1} \right) \\
&=e^{\ii\Sigma\phi/2} \ket{\Psi_\mathrm{s}} \otimes \left( e^{\ii\Delta\phi/2}\ket{0}+e^{-\ii\Delta\phi/2}\ket{1} \right) \nonumber
\end{align}
where the phases $\phi_i$ contain the stored magnetic field information (see Fig.~\ref{fig:1}b and Fig.~\ref{fig:app_gates}).
The initial phases are $\phi_{i=0,1}=0$.
Note that $\Sigma\phi = \phi_0+\phi_1$ constitutes a global phase and only the interference term $\Delta\phi=\phi_0-\phi_1$ is accessible at the final read out.
\\
\indent
During sensing steps, NOT-gates on the sensor spin conditional on the state of the memory spin (i.e. \cMnotS-gates) entangle and disentangle sensor and memory (see Fig.~\ref{fig:app_gates}a).
While sensor and memory are entangled, a phase $\varphi$ is accumulated, which linearly depends on sensing time $\tau$ and the local magnetic field, for example due to hyperfine coupled target spins.
The actual conditions of the \cMnotS-gates, their order and the initial sensor state determine to which storage phase $\phi_i$ the sensing phase $\varphi$ is added. 
During all three storage steps between the sensing steps, we manipulate target nuclear spins with radiofrequency (RF) pulses to induce a measurement signal.
Target spins that are flipped in all three storage periods contribute maximally to the accumulated signal, while the total phase due to quasi-static magnetic field noise is filtered out (i.e. does not contribute to $\Delta\phi$).
\\
\indent
Fig.~\ref{fig:2}a,b show exemplary NMR spectra of weakly coupled \Ciso\ target spins obtained with our measurement sequence as follows.
We set the total sensing time to $\tau=100$ and $200\,\mu$s in Fig.~\ref{fig:2}a and b, respectively.
During first and last storage time, we perform each one $\approx 40\,\mu$s-long RF $\pi$-pulse on all target spins within a spectral range of about $\Delta\nu=\pm10\,$kHz around the RF frequency (see Fig.~\ref{fig:app_gates}d).
Their main purpose here is to enable sensitivity to all the target spins despite the dynamical decoupling sequence of the sensor spin.
During the the correlation time \Tcorr\ however, we apply a long $\pi$-pulse ($430$ and $860\,\mu$s in Fig.~\ref{fig:2}a and b), which in turn is selective in a narrow spectral range of about $\pm1$ and $\pm0.5\,$kHz and therefore determines spectral resolution.
Only those target spins contribute maximally to the signal, which are flipped during all three storage steps.
\\
\indent
While the upper spectrum in Fig.~\ref{fig:2}a reveals one resonance at the bare \Ciso\ Larmor frequency, the lower spectrum and its zoom-in in Fig.~\ref{fig:2}b show multiple target spins shifted by their individual coupling \Azz.
To this end, we have adapted the condition of the \cMnotS gates during the sensing steps, such that the sensor is in its non-magnetic $\ket{0}$ or magnetic $\ket{1}$ state during \Tcorr.
The latter case switches on coupling and therefore the possibility to manipulate individual target spins conditional on their coupling.

\subsection{Composition of controlled sensing gates}
\label{sec:gate_composition}
In this section, the utilized conditional gates for the sensing steps shown in Fig.~\ref{fig:1} to \ref{fig:3} shall be explained in detail.
The sensing steps as used in this work, expect the sensor memory state to be of the form like in eq.~(\ref{eq:storage_state}) with $\ket{\Psi_s} = \ket{\mathrm{s_i}}$ prior to the sensing step.
The state after the sensing step is of the same shape with $\ket{\Psi_s} = \ket{\mathrm{s_f}}$ ($\mathrm{s_i,\,s_f} \in \{0,1\}$).
Sensing steps can therefore flip or not flip the sensor state, while adding a phase $\varphi$ to either phase $\phi_0$ or $\phi_1$ of the memory superposition state (see eq.~(\ref{eq:storage_state})).
We end up with four different gates, adding a phase to one of the two memory states, while flipping or not flipping the sensor spin (see Fig. \ref{fig:app_gates}a).
It can be seen, that two identical consecutive \cMnotS-gates separated by an evolution time \textbf{$\tau$} do not change the sensor spin state state (i.e. $\ket{\mathrm{s_f}} = \ket{\mathrm{s_i}}$), however accumulate a phase onto one of the memory spin states (see Fig.~\ref{fig:app_gates}a left column).
By using two \cMnotS-gates with different conditional states, the sensor state can be flipped in addition to the phase accumulation (i.e. $\ket{\mathrm{s_f}} = \ket{\mathrm{\overline{s}_i}}$, see Fig.~\ref{fig:app_gates}a right column).
Depending on the initial sensor state $\mathrm{s_i}$ and the conditional state of the first \cMnotS-gate of the pair, the phase $\varphi$ adds either to $\phi_0$ or $\phi_1$ (identical for rows of Fig.~\ref{fig:app_gates}a).
%

\subsection{Associated filter functions}
\label{sec:filter_functions}
The overall measurement sequences should constitute dynamical decoupling sequences similar to a spin-echo (SE), a Carr-Purcell-Meiboom-Gill (CPMG) or a periodic dynamical decoupling (PDD) sequence \cite{cywinski_how_2008}.
Therefore, the phases $\varphi$ accumulated during successive sensing steps should be added to opposite phases $\phi_0$ or $\phi_1$ to mimic the $\pi$-pulse effect of the standard dynamical decoupling sequences, which constantly switches the magnetic field sensitivity between 1 and -1.
Different from the standard techniques, our sequence of sensing steps is interleaved with storage times exhibiting no magnetic field sensitivity.
The magnetic field sensitivities at times $t^\prime$ during the run of a dynamical decoupling sequence of duration $t$ are given by the function $f(t,t^\prime)$.
Given the requirement of switching sensitivities and the option to set a certain sensor eigenstate during individual storage intervals the proper quantum gates for the series of sensing steps can be constructed according to section \ref{sec:gate_composition}.
Switching of sensitivities also requires flips of target spins during each storage period in order to be sensitive to their field exerted on the sensor.
\\
\indent
As for standard dynamical decoupling sequences, we can deduce a filter function $F(\omega t)$ of our full sequence, where $\omega$ the angular frequency of a potential oscillating magnetic field and $t$ is the total duration of a sequence.
Then, $F(\omega t) \omega^{-2}$ expresses the spectral sensitivity to magnetic field noise \cite{cywinski_how_2008,bar-gill_suppression_2012}.
The decay of coherent phase information $W(t)=\ee^{-\chi(t)}$ can then be expressed via 
\begin{equation}
\chi(t) = \pi^{-1} \int^\infty_0{d\omega S(\omega)\frac{F(\omega t)}{\omega^2}},
\label{eq:chi_decay}
\end{equation}
where $S(\omega)$ is the noise-spectral-density.
The filter function is obtained by Fourier-transforming the magnetic field sensitivity function $f(t,t')$ of the measurement sequence with respect to $t'$ yielding $\tilde{f}(t,\omega)$.
\begin{equation}
F(\omega t) = \frac{\omega^2}{2} \left| \tilde{f}(t,\omega) \right|^2
\label{eq:filter_function}
\end{equation}
\\
\indent
The filter functions for the measurement sequences performed here and for the slightly simpler version of Ref.~\onlinecite{zaiser_enhancing_2016-1} are
\begin{align}
\label{eq:filter_function_here}
F(\omega t) =& \, 32  \sin^2 \frac{\eta_\tau \omega t}{8}  \sin^2 \frac{\left(2-\eta_\tau-2 \eta_{T\mathrm{c}} \right) \omega t}{8} \nonumber \\
& \times \cos^2 \frac{\left(1+\eta_{T\mathrm{c}}\right)\omega t}{4}\\
\text{and} & \nonumber \\
F(\omega t) =& \, 8  \sin^2 \frac{\eta_\tau \omega t}{4}  \sin^2 \frac{\left(2-\eta_\tau \right) \omega t}{4} \text{,} 
\end{align}
respectively, where $\eta_\tau$ and $\eta_{T\mathrm{c}}$ are fractional sensing and central correlation times normalized by total sensing time $t$ and obeying $0\le \eta_\tau+\eta_{T\mathrm{c}},\, \eta_\tau \, , \eta_{T\mathrm{c}}\le 1$.
\\
\indent
Examples of both filter functions multiplied by $2/\omega^2$ are plotted in Fig.~\ref{fig:app_gates}e for sensing time $\tau=0.3\,$ms, duration of first and last storage interval each with RF $\pi$-pulse of $0.04\,$ms (where applicable) and variable central correlation time \Tcorr\ up to 5\,ms.
For comparison, the filter functions for Ramsey, SE and PDD sequence, each of equal total sensing time, are displayed.
On the left panel, the sequence of Ref.~\onlinecite{zaiser_enhancing_2016-1} is shown (i.e. $\frac{\tau}{2} - \Tcorr - \frac{\tau}{2}$) and on the right panel, the current sequence (i.e. $\frac{\tau}{4} - \pi_\mathrm{RF} - \frac{\tau}{4} - \Tcorr - \frac{\tau}{4} - \pi_\mathrm{RF} - \frac{\tau}{4}$).
Apparently, the first filter function resembles that of a spin echo for small correlation times.
However, for large $\Tcorr\gg\tau$ it becomes a fast oscillating function under the envelope of a Ramsey filter function.
Thus, it gets sensitive to small frequency noise.
The sequence used in the current paper circumvents this issue by adding another filter step into initial and final sensing interval.
For negligible storage times, the new sequence resembles a PDD3 sequence and for large \Tcorr\ it also shows fast oscillations under the envelope of a SE filter function.
Hence, sensitivity to small frequency noise is reduced.

\section{characterization of sensor and memory spin qubits}
\label{app:sensor_and_memory}
In the NV center hybrid sensor, the electron and nuclear spin have fundamentally different tasks because of their different properties.
While the electron spin is very susceptible to magnetic fields, the nuclear spin is almost unaffected by it.
Yet, the nuclear spin is quite strongly coupled to the electron spin when compared to present relaxation rates.
Hence, the electron spin serves as the primary transducer from magnetic fields to a quantum phase (sensor) whereas the nuclear spin is ideally suited for storage of the latter phase (memory).
Here, we briefly characterize both spins under the current sample and setup conditions (see Appendix~\ref{app:setup}). 
The spin Hamiltonian describing the combined sensor memory and the target spins is
\begin{align}
	\label{eq:H_tot}
	H =& H\sensorindex + H\memoryindex + H\sampleindex + H\couplingindex \\
	  =& DS_z^2 + \tilde{\gamma}\sensorindex B_z S_z + \tilde{\gamma}\memoryindex B_z I_z\memoryindex \nonumber \\
		&+ \tilde{\gamma}\sampleindex B_z \sum_\mathrm{samp}{ I_{z}\sampleindex } + S_z A_\parallel\memoryindex I_z\memoryindex \nonumber \\
		&+ S_z \sum_\mathrm{samp}{A_\parallel\sampleindex I_z\sampleindex} \, .\nonumber
\end{align}
In equation (\ref{eq:H_tot}) the spin operator for the sensor is described by $S_z$ and the memory and target spin operators by $I_z\memoryindex$ and $I_z\sampleindex$ respectively, where the $z$-axis coincides with the static magnetic field direction and the NV center symmetry axis.
Furthermore, the gyromagnetic ratios divided by $2\pi$ are given by the respective $\tilde{\gamma}$ and the crystal-field splitting of the NV sensor spin triplet ($S=1$) is denoted by $D$.
Finally, we account for the hyperfine coupling of the sensor to the memory and target spins only via the respective longitudinal coupling constants $A_\parallel$.
The latter are $A_\parallel\memoryindex=-2.16\,$MHz for the memory and around $A_\parallel\sampleindex\sim1\,$kHz for the target spins.
The latter follows from the secular approximation and the fact that the symmetry axis of the \Niso\ hyperfine tensor is collinear with the $z$-axis and couplings to target spins are expected to be much smaller than their nuclear Zeeman energies.

\begin{figure}[t!]
	\begin{center}
	\includegraphics[width=\columnwidth]{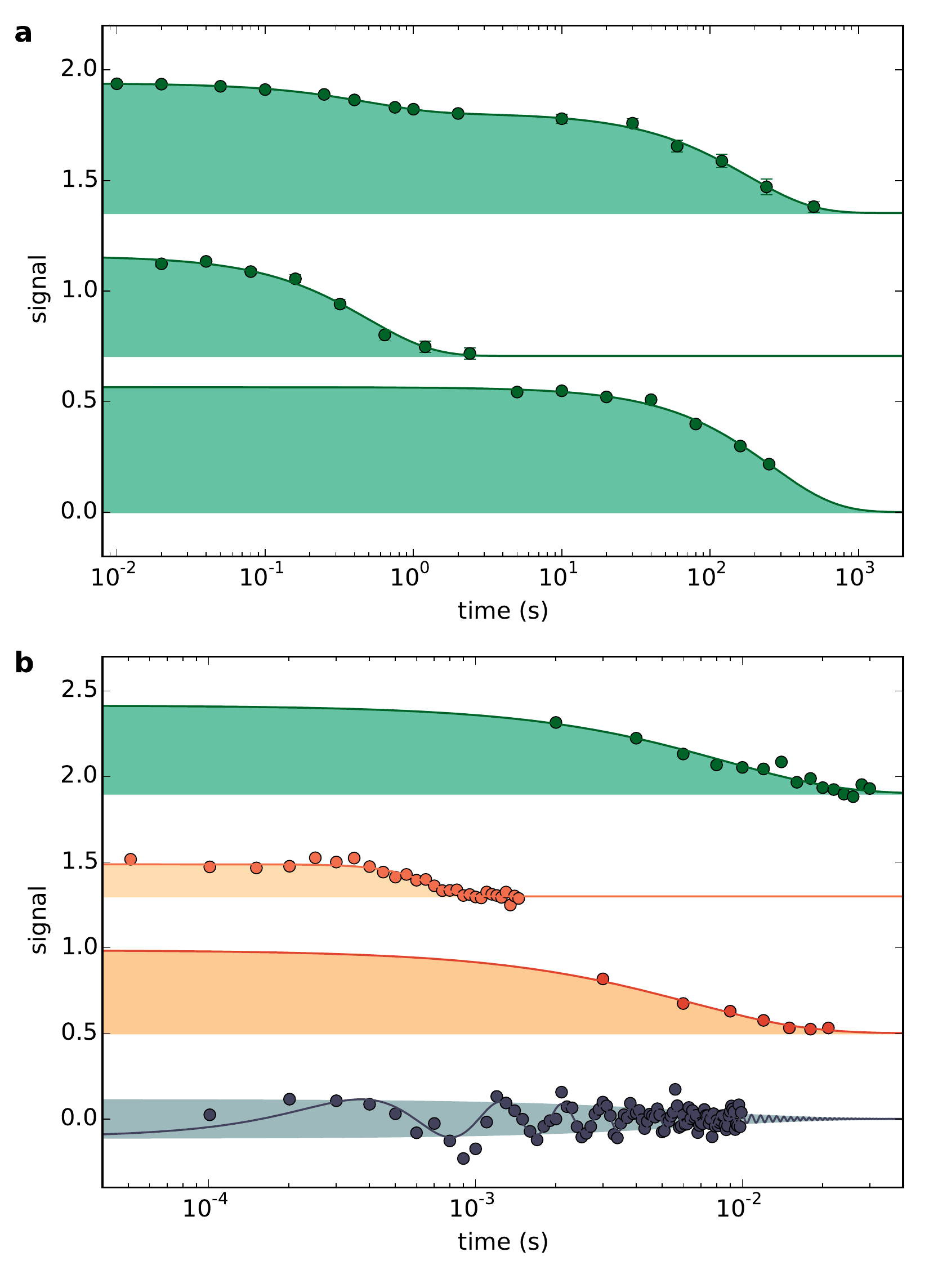}
	\caption{
		\textbf{decay constants of sensor memory and target}.
		\textbf{(a)} shows the lifetime of the memory spin.
		The uppermost graph shows the \Tonememory lifetime.
		There are two different decays visible, one with a decay constant of $0.49\,$s, the other $180\,$s.
		The fast decay corresponds to the measurement runs, during which the NV center resides in the \NVz charge state, the slow decay corresponds to the \NVm case.
		By using charge state readout (see ref.~\onlinecite{aslam_photo-induced_2013}) and data post selection, one can isolate the pure \NVz and \NVm measurements.
		The result can be seen in the middle (\NVz) and lower (\NVm) line.
		\textbf{(b)} shows the sensor performance, including the coherence of the probed spins, without any dissipative decoupling.
		The uppermost line shows the coherence decay of the memory spin, with a decay constant of \Ttwomemory$=(8.6\pm1.3)\,$ms.
		Below, the orange lines show from top to bottom the coherence (\Ttwosensor\mbox{=$688\pm31\,\mu$s}) and longitudinal relaxation (\Tonesensor $=6.4\pm0.6\,$ms) of the sensor spin.
		The black line shows the coherence of the target spins, with a decay constant of \ttwostar$=6.78\pm1.3\,$ms.
		Please note, that the \Tonesensor decay limits \Ttwomemory, as well as \Ttwostar (see sec.~\ref{sec:master_equation}).
		\label{fig:app_C_1}
	}
	\end{center}
\end{figure}

\subsection{The electron spin sensor}
The NV electron spin has a longitudinal relaxation time constant of $\Tonesensor\approx 6\,$ms, which is in good agreement with literature values \cite{jarmola_temperature-_2012}.
We attribute \Tonesensor\ to the time constant of polarization decay in $m_S=0$ and appearance in $m_S=\pm1$ after initialization into $m_S=0$.
The difference of both signals is shown in the lower orange line in Fig.~\ref{fig:app_C_1}b.
The decay of the NV electron spin polarization limits the lifetime of coherence of any superposition state of this electron spin.
In addition, all other spins in the vicinity of the electron spin, which usually couple like $\propto S_zI_z$, will experience decoherence.
\\
\indent
The coherence lifetime of the electron spin superposition state $\ket{0}+\ket{-1}$ is measured to be $\Ttwosensor=688\,\mu s$ via a spin echo measurement (see green line in Fig.~\ref{fig:app_C_1}b).
This value determines the limit for coherent phase accumulation in any measurement scenario.
It does, however, not limit any storage time of phases.
Rather, the coherence lifetime limits the access to strongly coupled nuclear spins (i.e. coupling $>1/\Ttwosensor$).

\subsection{The nuclear spin memory}
The \Niso\ nuclear spin is very isolated in the diamond lattice in a sense that no other nitrogen spin (i.e. $^{14}$N) is available.
The \Niso\ spin is most strongly coupled to the NV electron spin via the hyperfine interaction $H_{hf}=\underline{S}\mathbf{A}\underline{I}$ with tensor $\mathbf{A}$.
Coupling to other electron or nuclear spins is negligible.
In Hamiltonian (\ref{eq:H_tot}) we have neglected all parts of the hyperfine tensor except for \Azz.

\subsubsection{The memory's classical storage time}
The memory spin's \Tonememory\ lifetime is measured for three different settings, namely, for NV$^-$ under dark conditions, for NV$^0$ in the dark and for the case of continuous single shot readout of the memory spin state \cite{neumann_single-shot_2010,waldherr_quantum_2014,zaiser_enhancing_2016-1}.
The memory $\Tonememory$ lifetime during single shot readout scales quadratically with increasing magnetic field \cite{neumann_single-shot_2010} and reaches a value of around 1\,s at 3\,T (see Fig.~\ref{fig:1}c).
The lifetime during single shot readout depends on the actual timing of MW, wait and laser pulses during readout.
Therefore, we also provide the lifetime in terms of laser pulses on the right vertical axis for comparison.
The quadratic magnetic field scaling of the $\Tonememory$ lifetime also holds for \NVz\ in the dark, it reaches around 2\,s at 3\,T (see Fig.~\ref{fig:1}c).
\\
\indent
The scaling is different for the negative NV center in the dark (see Fig.~\ref{fig:1}c).
There, the $\Tonememory$ lifetime increases first but then levels off at a value of $280\,$s at 1.5\,T.
Nevertheless, $280\,$s and $2\,$s for \NVm\ and \NVz, respectively, yield quite useful storage times for classical information.
\\
\indent
In the following we describe the measurement of $\Tonememory$ in the dark for the two charge states \NVm and \NVz.
In a first measurement we did not discriminate between the charge states.
We perform a single \Niso\ spin readout step with a duration of a few ms, which yields either low fluorescence for $m_I=+1$ or high fluorescence for $m_I=0,-1$.
We  repeat these readout steps until the outcome is ``low''.
In the latter case we switch off the laser illumination via the AOM for a time $\tau$.
For $\tau$ larger than 5\,s, we additionally flip a beam block into the laser path to avoid laser leakage through the AOM.
After time $\tau$ we perform another readout step and record the result.
Since we did not discriminate the charge states, we do see two decay processes in the upper curve of Fig.~\ref{fig:app_C_1}a.
\\
\indent
In a second measurement we do discriminate between charge states.
Therefore, we add a charge state detection sequence before and after step (3).
To this end, we send a weak $1\,\mu\text{W}$ orange ($594\,$nm) laser beam for 20\,ms and record the fluorescence.
In case of \NVz\ almost only dark counts are detected and in case of \NVm\ the photon number is considerably higher and thus the charge states can be discriminated \cite{aslam_photo-induced_2013}.
This way we can sort events of \NVz\ and \NVm\ lifetime measurements (see the two lower curves in Fig.~\ref{fig:app_C_1}a).
In addition, we can check whether the charge state changes during the dark period.
We did not see any hint for the latter charge state changes in our measurements on the given timescale.
\\
\indent
In the following we consider reasons for the \Niso\ \Tonememory\ measurement results.
The longitudinal relaxation is not affected by mutual cross-relaxation with other nitrogen spins due to absence of the latter.
Minor relaxation rates arise from spin state mixing with the sensor spin via the transverse parts of the hyperfine coupling tensor $\mathbf{A}$ (i.e. $A_{\perp}\sim4\,$MHz).
The mixing $\epsilon$ ($\ket{m_S,m_I}=\sqrt{1-\epsilon}\ket{-1,0}+\sqrt{\epsilon}\ket{0,-1}$) is suppressed by the strong detuning of sensor and memory spin resonances due to the sensor Zeeman energy ($E_\mathrm{eZ}\sim 40\,$GHz), i.e. $\epsilon\sim 10^{-8}\propto (A_{\perp}/E_\mathrm{eZ})^2$.
The mixing in combination with sensor spin projection event into an $m_S$ eigenstate leads to a flip of the memory spin with probability $\epsilon$.
\\
\indent
If the NV center resides in the dark and in its negative charge state, the sensor spin flips with a rate of $\sim 1\,$kHz.
Therefore, the expected limit on the memory spin $\Tonememory$ lifetime is on the order of $10^5$s.
In the neutral charge state, the electron spin flip rate is on the order of $\sim 1\,$MHz and thus we expected memory spin $\Tonememory\sim 100\,$s.
During single shot readout of the memory spin, frequent optical excitations lead to stronger mixing because of $A_{\perp}\approx 40\,$MHz in the excited state.
In addition, the sensor spin is projected much more often ($\sim 1\,$MHz).
Thus the \Niso\ lifetime during single shot readout is expected to be $\sim 1\,$s.
\\
\indent
Whereas the memory spin $\Tonememory$ lifetime during single shot readout is in agreement with our estimates, the lifetimes under dark conditions and in the negative and neutral charge state are much smaller than expected (see Fig.~\ref{fig:1}c), however the expected order of lifetimes for the different conditions is reproduced.
One yet unconsidered reason for \Niso\ $\Tonememory$ relaxation processes might be the quadrupole moment of the nucleus, which couples to electric field gradients.

\subsubsection{The memory's quantum storage time}

Because of the particularly strong coupling of the memory to the sensor compared to that of the target spins, the $T_1$ influence of the sensor spin can not be suppressed in our experimental scheme.
\\
\indent
Theoretically, the coherence decay constant \Ttwomemory\ 
is expected to be $3/2\cdot \Tonesensor$ 
if only affected by the sensor spin \Tonesensor\ 
(see Appendix~\ref{app:relaxation_theory}).
Indeed we obtain 
$\Ttwomemory=8.6\pm1.4\,$ms, which is longer than \Tonesensor\ 
(
$\Ttwomemory=1.35 \, \Tonesensor$ 
).
Similar values have been obtained in Ref.~\onlinecite{waldherr_high-dynamic-range_2011}.
Other influences that might lead to a deviation from the theoretical expectation are laser leakage accompanied by additional sensor spin decay rates into $m_S=0$.
\\
\indent
Here, the measurement of \Ttwomemory\ 
is performed via a correlation measurement as discussed in Ref.~\onlinecite{zaiser_enhancing_2016-1}.
Hence, we have created an equal superposition state on the memory.
Then, we have correlated the phase of the memory with the current magnetic field via two sensor CNOT gates separated by $\tau=1\,\mu$s.
The following free evolution time was swept and is displayed as horizontal axis in Fig.~\ref{fig:app_C_1}b.
Finally, another pair of sensor CNOT gates separated by $\tau$ is applied, before the phase of the memory is read out (see green line figure~\ref{fig:app_C_1}b).
The result demonstrates that also coherent metrology information can be stored beyond \Tonesensor. 

\section{Theoretical derivation of sensor relaxation effects on memory and target spins}
\label{app:relaxation_theory}
In this section we are going to derive the effect of sensor relaxation on memory and target spins using a master equation approach.

\subsection{Master equation for sensor spin in the dark}
\label{sec:master_equation}
According to eq.~(\ref{eq:H_tot}) backaction from the sensor spin on memory and target spins is only mediated via a $A_\parallel S_zI_z$ coupling term.
Thus stochastic flips of the sensor spin (i.e. \Tonesensor\ decay) will lead to decoherence of memory and target spins (i.e. \ttwostar\ and \Ttwomemory\  decay) due to an unknown phase accumulation.
\\
\indent
Here, we concentrate on the situation of device and target spins in the dark.
Relaxation effects can be modeled by investigating a master equation for the combined system of sensor and memory spin with density matrix $\rho=\sum_k{\rho_{\mathrm{e},k} \otimes \rho_{\mathrm{n},k}}$,
\begin{align}
\label{eq:master}
\dot{\rho}(t) =&-i2\pi[H,\rho(t)] \\
&+ \sum_j{ L_j \rho L_j^\dagger - \frac{1}{2} \left( L_j^\dagger L_j \rho + \rho L_j^\dagger L_j \right) } \, ,\nonumber
\end{align}
with the Lindblad operators $L_j$ describing the stochastic flips of the sensor spin.
Given the quasi infinite temperature in the current experiment (i.e. $E/k_\mathrm{B}T=0.014$) the latter decay can be modeled via the a depolarizing quantum operation $\mathcal{E}(\rho_\mathrm{e}) =\left( 1-\frac{\Delta t}{\Tonesensor} \right) \rho_\mathrm{e} + \frac{\Delta t}{\Tonesensor} \frac{\mathbf{1}_\mathrm{e}}{3}$ where $\rho_\mathrm{e}$ is the sensor spin density operator.
An exemplary operator sum representation is
\begin{align}
\label{eq:quantum_op}
\mathcal{E}(\rho_\mathrm{e}) &= \sum_k{E_k \rho_\mathrm{e} E_k^{\dagger}} \\
&= \left( 1-\frac{\Delta t}{\Tonesensor} \right) \rho_\mathrm{e} + \frac{\Delta t}{3 \Tonesensor} \sum_{n,m=-1}^{1}{ \ketbra{n}{m} \rho_\mathrm{e} \ketbra{m}{n} } \nonumber
\end{align}
describing a time-step $\Delta t \ll \Tonesensor$ and obeying $\sum_k{E_k^\dagger E_k}=\mathbf{1}$.
From eq.~(\ref{eq:quantum_op}) we deduce the Lindblad operators as $L_{j=3m+n+5} = \frac{1}{\sqrt{3 \Tonesensor}} \ketbra{m}{n} \otimes \mathbf{1}_\mathrm{n}$ with $m$ and $n$ being the sensor spin projections  $m_S =-1,0,1 $ and hence $j=1,..,9$.
This quantum operation reproduces the $\Tonesensor$ relaxation of the sensor spin state polarization.
The concomitant sensor spin decoherence is irrelevant in what follows and therefore does not need to be reproduced.
The decoherence is anyway also affected by other sources such as paramagnetic impurities in diamond or the nuclear spin bath.
\\
\indent
Next, we solve eq.~(\ref{eq:master}) numerically with initial state $\rho(0)=\ketbra{0}{0} \otimes \ketbra{x}{x}$ and observe the remaining coherence via $\mathrm{Tr} \left\{ \mathrm{Tr_e} \left[ \rho(t) \right] \sigma_x\right\}$.
When varying the coupling term $A_\parallel$ we observe two different regimes (see Fig.~\ref{fig:4}d).
For $A_\parallel > 1/\Tonesensor$ (e.g. as for the memory qubit) the nuclear spin coherence time $\ttwostar=3 \Tonesensor/2$, and for $A_\parallel \ll 1/\Tonesensor$ (e.g. as for weakly coupled target spins) $\ttwostar$ grows as $A_\parallel^{-2}$.
\\
\indent
The first regime (i.e. $A_\parallel > 1/\Tonesensor$) can be explained by considering the action of the depolarizing quantum operation for initial state $m_S=0$.
The initial decay out of $m_S=0$ happens with rate $2 /3\Tonesensor$ whereas the steady state $\mathbf{1}_\mathrm{e}$ is reached with rate $1/\Tonesensor$.
For couplings $A_\parallel > 1/\Tonesensor$ nuclear spin decoherence happens instantly (i.e. in $\Delta t\ll \Tonesensor$) upon a transition from $m_S=0$ into $m_S=\pm1$.
Hence, the amount of nuclear spin coherence is a measure for the probability that not a single sensor spin flip has yet occurred whereas the population of $m_S=0$ is influenced by rates from and towards $m_S=0$.
\\
\indent
For the second regime (i.e. $A_\parallel \Tonesensor \ll 1$) the sensor spin adds tiny additional random phases $\delta \phi \sim A_\parallel \Tonesensor m_S\ll1$ to a nuclear spin superposition state during subsequent sensor spin flips.
The phase uncertainty grows with time $t$ as $\sigma_\phi \propto \delta \phi \sqrt{t/\Tonesensor}$.
Hence, the dephasing time $\ttwostar$ of the nuclear spin scales as $\ttwostar \propto \Tonesensor/\delta \phi^2 \propto (\Tonesensor)^{-1}A_\parallel ^{-2}$.
\\
\indent
For the NV center in the negative charge state (\NVm) we obtain a nuclear spin dephasing time of around $9\,$ms for couplings down to $A_\parallel \sim 50\,$Hz, which corresponds to a sensor-proton-spin-distance of $\sim 15\,$nm (see Fig.~\ref{fig:4}d).
For smaller couplings the nuclear spin $\ttwostar$ shows the quadratic increase with inverse coupling constant as derived above.
\\
\indent
For the NV center in the neutral charge state (\NVz) the setting is different.
In its ground state it features an orbital and an electron spin doublet \cite{gali_theory_2009-1}.
Previous experiments revealed fast decoherence of the \Niso\ memory spin under these circumstances \cite{waldherr_dark_2011}, which can be modeled by an electron spin $S=1/2$ with a lifetime $T_1\nvzindex\approx 13\,\mu$s.
Given $T_1\nvzindex$ we can simulate the effect on the $\ttwostar$ coherence lifetimes of nuclear spins with varying coupling.
The resulting behavior is displayed as the red line in Fig.~\ref{fig:4}d.
For couplings stronger than $\approx50\,$kHz the nuclear spin lifetime is limited to $\approx 20\,\mu$s and for smaller couplings we see the mentioned quadratic increase.
Interestingly, for couplings smaller than $\sim1.7\,$kHz nuclear spin $\ttwostar$ lifetimes for the \NVz case do overtake the lifetimes for the \NVm case.

\subsection{Increased dissipation by illumination}\label{sec:increased-dissipation-by-illumination}
\begin{figure}[t!]
	\begin{center}
\includegraphics[width=\columnwidth]{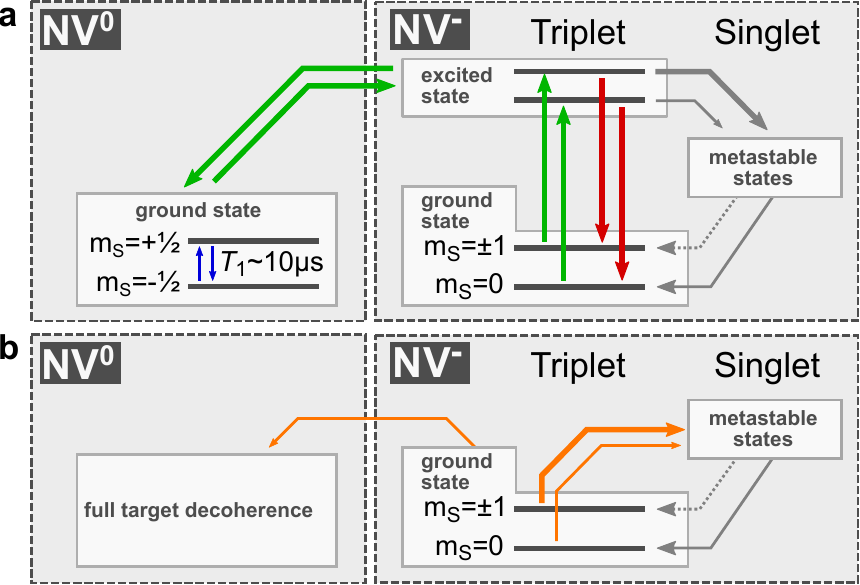}
	\caption{
		\textbf{NV level schemes for Simulation}.
		\textbf{(a)} Level scheme showing the important parts of negative and the neutral charge state of the NV center.
		In NV-, the two spin sub-levels show different fluorescence intensities, as well as a spin polarization by an inter system crossing (ISC) into the metastable singlet state (here drawn as a "`black box"').
		Ionization occurs as an effective two photon process from the \NVm\ ground state \cite{aslam_photo-induced_2013}, ending up in the uncharged \NVz\, with a high spin dissipation rate.
		Recombination also occurs as a two photon process.
		For the simulation of the pure \NVm\ and \NVz\ case, only the respective ground states are modeled.
		\textbf{(b)} shows the reduced level scheme for the simulation of the continuous excitation case.
		The \NVm\ excited state is omitted due to its short lifetime. The spin dependent ISC is realized by spin dependent excitation rates.
		Ionization to the \NVz\ charge state is implemented by immediate target spin decoherence.
		\label{fig:app_simulation_levels}
	}
	\end{center}
\end{figure}

Apart from investigating the nuclear spin coherence for both charge states separately, we are also interested in the behaviour during illumination.
To model the illumination behavior of the \NVm\ charge state, we add an additional metastable state to the electronic level system.
Illumination and reinitialization of the electron spin is facilitated by spin state dependent rates into the metastable state, as well as different decay rates back into the different spin states.
However, illumination also opens up the door towards ionization of the NV center into the \NVz\ charge state.
Since \NVz\ can have deleterious effects on our classical memory, we want to investigate the case of continuous repolarization without ionization.
Ionization events are therefore modeled as instantaneous decoherence on our target spins (see Fig.~\ref{fig:app_simulation_levels}).
The ionization rates $\gamma_\mathrm{ion}$ depend quadratically on the excitation rate $\gamma_\mathrm{exc}$ (since it is a two photon process, see ref.~\onlinecite{aslam_photo-induced_2013}).
\begin{align}
	\label{eq:gamma_exc}
	\gamma _{\text{exc}}&=P_{\text{Laser}}\cdot c_\text{exc} \\
	\label{eq:gamma_ion}
	\gamma _\text{ion}&=(\gamma_\text{exc})^2\cdot c_\text{ion}
\end{align}
\\
\indent
We can numerically model the experimentally obtained $\ttwostar$ times for increasing laser power and constant coupling.
To this end, we have to adjust parameter such as $c_\text{exc}$ and $c_\text{ion}$, as well as $p_{\text{branching}}$ the probability of the metastable state to decay into the $m_s=0$ sublevel, to fit the simulated results to the experimental data (see Fig.~\ref{fig:3}a).
We obtain the values $c_\text{exc}=2.5\cdot10^5\frac{\text{excitations/s}}{\mu\text{W}}$, $c_\text{ion}=1/(7\cdot 10^8) \frac{\text{ionisations}}{\text{excitations}^2}$ and the branching ratio $p_{\text{branching}}=0.96$
We observe a tiny initial decrease in $\ttwostar$ time, where the increase in optically induced dissipation rate is smaller than the coupling.
When increasing the excitation rate beyond the coupling strength, we see an increase due to a sufficiently high and increasing dissipation rate. 
The $\ttwostar$ increase continues towards a global maximum of that model.
For higher excitation rates, ionization events destroy access to the memory.
\\
\indent
With the optimized model, we can extrapolate the optimum $\ttwostar$ time for decreasing target spin coupling strength.
In Fig.~\ref{fig:4}d we see the resulting increase of $\ttwostar$ with a slightly smaller slope as for the \NVz\ and \NVm\ case in the dark.
Of course, the model fits to our measurement results in the kHz coupling range.
Next, we check for consistency of the further increase.
\\
\indent
As in the dark cases, the coherence time scales as $\ttwostar\propto\Gamma/\Azz^2$ (cf. eq.~\ref{eq:motional_average}), where $\Gamma \propto \gamma_\mathrm{exc}$ (see eq.~\ref{eq:gamma_exc}).
However, at the optimum, \ttwostar\ is equally limited by the ionization rate as $\ttwostar\propto1/\gamma_\mathrm{ion}\propto 1/\Gamma^2$ (cf. eq.~\ref{eq:gamma_ion}), and therefore scales quadratically with the inverse optical excitation rate, as with the dissipation rate.
Hence, we require $\Gamma/\Azz^2\propto 1/\Gamma^2$ (i.e. $\Azz^2\propto\Gamma^3$) leading to a coherence time scaling of $\ttwostar \propto \Azz^{-4/3}$.
This result perfectly agrees with the numerically simulated scaling in Fig.~\ref{fig:4}d.

\begin{acknowledgments}
We thank Sebastian Zaiser, Ville Bergholm, Nikolas Abt, Ingmar Jakobi, Fedor Jelezko, Liam McGuinness and Johannes Greiner for fruitful discussions and technical advice.
We acknowledge financial support by the German Science Foundation (SFB-TR 21, SPP1601), the EU (DIADEMS), the Volkswagen Stiftung, the JST and JSPS KAKENHI (No.
26246001, 26220903).
\end{acknowledgments}

%

\end{document}

%% file: macros.tex
\usepackage{color}
\definecolor{dred}{rgb}{.9,0.2,.2}
\definecolor{ddred}{rgb}{.8,0.65,.65}
\definecolor{dblue}{rgb}{.2,0.2,1.0}
\definecolor{dgreen}{rgb}{.2,0.7,.2}

\newcommand{\out}[1]{}

\newcommand{\comment}[1]{}

\newcommand{\todo}[1]{}

\newcommand{\proton}{$^{1}$H}
\newcommand{\Ciso}{$^{13}$C}
\newcommand{\Cis}{$^{12}$C}
\newcommand{\Siso}{$^{29}$Si}

\newcommand{\Niso}{$^{14}$N}
\newcommand{\Nis}{$^{15}$N}
\newcommand{\NVm}{NV$^-$}
\newcommand{\NVz}{NV$^0$}

\newcommand{\ii}{\ensuremath{\mathrm{i}}}
\newcommand{\ee}{\ensuremath{\mathrm{e}}}

\newcommand{\ttwostar}{\ensuremath{T_{2}^{*}}}
\newcommand{\Ttwostar}{\ttwostar}

\newcommand{\sensorindex}{\ensuremath{^{\text{sens}}}}
\newcommand{\memoryindex}{\ensuremath{^{\text{mem}}}}
\newcommand{\sampleindex}{\ensuremath{^{\text{tar}}}}
\newcommand{\nvzindex}{\ensuremath{^{\text{NV0}}}}
\newcommand{\couplingindex}{\ensuremath{^{\text{coupl}}}}

\newcommand{\Tonesensor}{\ensuremath{T_{1}\sensorindex}}
\newcommand{\Ttwosensor}{\ensuremath{T_{2}\sensorindex}}
\newcommand{\Ttwomemory}{\ensuremath{T_{2}\memoryindex}}
\newcommand{\Tonememory}{\ensuremath{T_{1}\memoryindex}}
\newcommand{\Tcorr}{\ensuremath{T_\mathrm{c}}}
\newcommand{\Azz}{\ensuremath{A_{\parallel}}}

\newcommand{\cMnotS}{$\text{C}_{\text{m}}\text{NOT}_{\text{s}}$}

\newcommand{\ket}[1]{\ensuremath{\left| #1 \right \rangle}}

\newcommand{\ketbra}[2]{\ensuremath{ \left| #1 \right \rangle \! \left \langle #2 \right | } }